\documentstyle[12pt,epsfig,psfig]{article}
\begin{document}
\begin{center}

{\Large \bf Muon pair production by photons in atmosphere: Is any 
room left for high-energy muon astronomy?}

\baselineskip=24pt
\vspace{1cm}

{\large V. A. Kudryavtsev}

\vspace{0.5cm}
{\it Department of Physics and Astronomy, University of Sheffield, 
Sheffield S3 7RH, UK}

\vspace{1cm}
{\large \bf Abstract}
\end{center}

Production of muon pairs by high-energy photons in electromagnetic
and hadronic showers in atmosphere has been calculated. The
effect of muon pair production in hadronic Extensive Air Showers 
(EAS) is unlikely to be seen by next generation detectors.
Applications of muon pair production process in electromagnetic
showers to the very high energy gamma-ray astronomy is discussed.
It is shown that, although this process dominates over conventional
pion and kaon decay above a few TeV in photon-initiated showers and provides 
a distinctive signature of photon-induced event (muon pair), 
it is practically impossible to discriminate such events
statistically from the background of muon pairs produced in the
hadronic EAS. The rate of events is very low and requires
detectors of a huge size.

\vspace{0.5cm}


\vspace{1cm}
\noindent Corresponding author: V. A. Kudryavtsev, Department of Physics 
and Astronomy, University of 
Sheffield, Hicks Building, Hounsfield Rd., Sheffield S3 7RH, UK

\noindent Tel: +44 (0)114 2224531; Fax: +44 (0)114 2728079; 

\noindent E-mail: v.kudryavtsev@sheffield.ac.uk

\pagebreak

\noindent {\large \bf 1.  Introduction}
\vspace{0.5cm}

Muon pair production in
electromagnetic showers in atmosphere has been considered 
in the middle of 1980s
by Kudryavtsev and Ryazhskaya \cite{vak1} and Stanev et al.
\cite{stanev1} in connection with observed muon excess from Cyg X-3 
\cite{nusex,soudan}. 
Stanev \cite{stanev2} and Berezinsky et al. \cite{vs1} discussed
high-energy gamma-ray astronomy using underground detectors.
Recently muon pair production in hadronic Extensive Air Showers
(EAS) has been calculated \cite{vak2}.

Underground muon data from the MACRO \cite{macro} and LVD \cite{lvd}
experiments did not reveal any excess from any region of the sky
or any source detected at TeV energies by Cherenkov telescopes.
In fact, the sensitivity of the existing underground detectors
is not enough for such a detection assuming known
processes of muon production with known cross-sections used in 
the calculations mentioned above. However, only
experimental search for possible excess with underground
detectors
can really provide information about processes of acceleration at
sources, muon production mechanisms and possibly give some indications
about new physics.

High-energy gamma-ray astronomy with muons was 
discussed also in a few recent papers \cite{halzen,bhatt}.
Halzen et al. \cite{halzen} considered photoproduction
of muons below 1 TeV and found that future 1 km$^3$ underwater/under-ice
neutrino telescopes (see, for example, \cite{amanda,antares}) 
can be able to detect muons associated with
high-energy photons from point sources. Bhattacharyya \cite{bhatt}
calculated gamma-induced muon flux from the Crab Nebula and obtained
a value within a reach of future neutrino telescopes.

Previous calculations, however, have been performed using 
one-dimensional model which restricted the analysis of
the detector sensitivity. No muon propagation in rock or water
was included in the calculations so far. 
At present powerful three-dimensional 
Monte Carlo codes for development of the showers in the
atmosphere (for example, CORSIKA \cite{CORSIKA}) make
possible full simulation of electromagnetic, hadronic and
muonic components of EAS produced by any primary particle.
Muons can be propagated through the rock or water down to
observation level using three-dimensional transport codes.
This allows calculation of detector sensitivity to
photon-induced muons taking into account effects of
muon scattering in atmosphere and rock.

In this paper we present new calculations of muon
production in photon-induced showers in atmosphere using
three-dimensional simulations. We also used
three-dimensional transport of muons through rock and water
to estimate sensitivity of existing and future detectors.
We are interested primarily in high-energy muons ($E_{\mu} 
\ge 100$ GeV) capable of reaching deep underground/underwater
experimental facilities.
We concentrate mainly
on direct muon pair production by photons because this
process can give a distinctive signature of gamma-induced
event (muon pair with small separation). We also present
calculations of muon pair production in hadronic showers.
The paper is organised in the following way. In Section 2
we describe method of calculation. In Section 3 we show
the results for muon pair production in hadronic showers
and compare them with previous calculations. Section 4
contains the results for muon production in electromagnetic
showers and comparison with earlier calculations. 
We discuss applications of our results for gamma-ray
astronomy in Section 5. The conclusions are given in Section 6.

\pagebreak

\vspace{0.7cm}
\noindent {\large \bf 2. Method of calculations}
\vspace{0.5cm}

Extensive Air Showers (electromagnetic and hadronic) at
vertical were
simulated using CORSIKA Monte Carlo code \cite{CORSIKA}.
Hadronic showers were produced by vertically incident 
protons with power-law spectrum with index $\gamma=2.7$.
For evaluation of muon intensities and yields we used
differential primary spectrum of all nucleons in the form:
$1.8 \times E^{-2.7}$ nucleons/(cm$^2$ s sr GeV)$^{-1}$
\cite{gaisser}. For a given energy per nucleon
heavy primaries will produce more muons and
photons than proton primaries. Since 
the number of muons and photons in a shower is proportional 
to the atomic weight of primary nucleus (in accordance with
our simulations), our approach to use all-nucleon spectrum
is justified regardless of the primary cosmic-ray composition.

Primary photons were assumed to have energy spectrum
with power index $\gamma=2.1$ and $\gamma=2.5$. Two 
different assumptions were made about high-energy cut-off.
In the first optimistic case no cut-off was included in the 
simulations. This may correspond to the nearby gamma-sources
with accelerated protons producing high-energy photons.
In the second, more realistic, case high-energy cut-off was 
assumed to be equal to 100 TeV. 

Muon pair production by photons
is included in the complete version of the CORSIKA code. In an ideal
case most important processes of muon production (conventional
pion and kaon production and decay, and direct muon pair
production) can be simulated in CORSIKA. In practice, 
we need many thousands or even many ten thousands muons 
at surface to estimate detector sensitivity underground. Such
a statistics can be easily obtained for conventional muon
production in hadronic showers. The probability of muon
production in electromagnetic showers is at least one order of 
magnitude less. This requires much higher statistics
for simulated electromagnetic showers. To overcome this
difficulty we used CORSIKA to simulate conventional
muons (from pion and kaon decay) and to obtain average 
atmospheric depth profiles
of photons with various energy thresholds in electromagnetic 
and hadronic showers.
Average depth profiles of photon fluxes
from CORSIKA output were then used to simulate muon pair 
production by photons. Thus, we separated muon pair production
by photons from the development of the shower. The advantage of
this is the dramatic reduction in the CPU time needed to
obtain sufficient statistics for underground muons. The CPU
consumption is driven by the statistics for photons in 
atmospheric showers. The disadvantage is the impossibility to 
consider two muon pairs
or muon pair and muons from pion decay in the same shower.
Note, however, that the probability of muon production in an
electromagnetic shower is quite small and is certainly much less
than 1 for all muon and photon energies important for further
considerations (only values of $x$ within the region
$10^{-3}<x=E_{\mu}/E_{\gamma}<1$, where $E_{\mu}$
is the muon energy and $E_{\gamma}$ is the initial photon energy,
contribute to muon production). This means that it is unlikely
to have two direct muon pairs or direct muon pair and conventional 
muons in the same shower. Further muon propagation through the 
rock/water also suppresses muon multiplicity.
As we will show later on, 
at all energies the integral probability of direct muon pair production
(muon yield) is less than $10^{-3}$ in an electromagnetic shower 
and, then the probability 
to have two muon pairs in the same shower is on average $10^{-3}$ times less.
At low muon energies the conventional muon yield may be quite high
but again the small probability of direct muon pair production
does not add much to the resulting muon flux.

\begin{figure}[htb] 
\begin{center}
\epsfig{figure=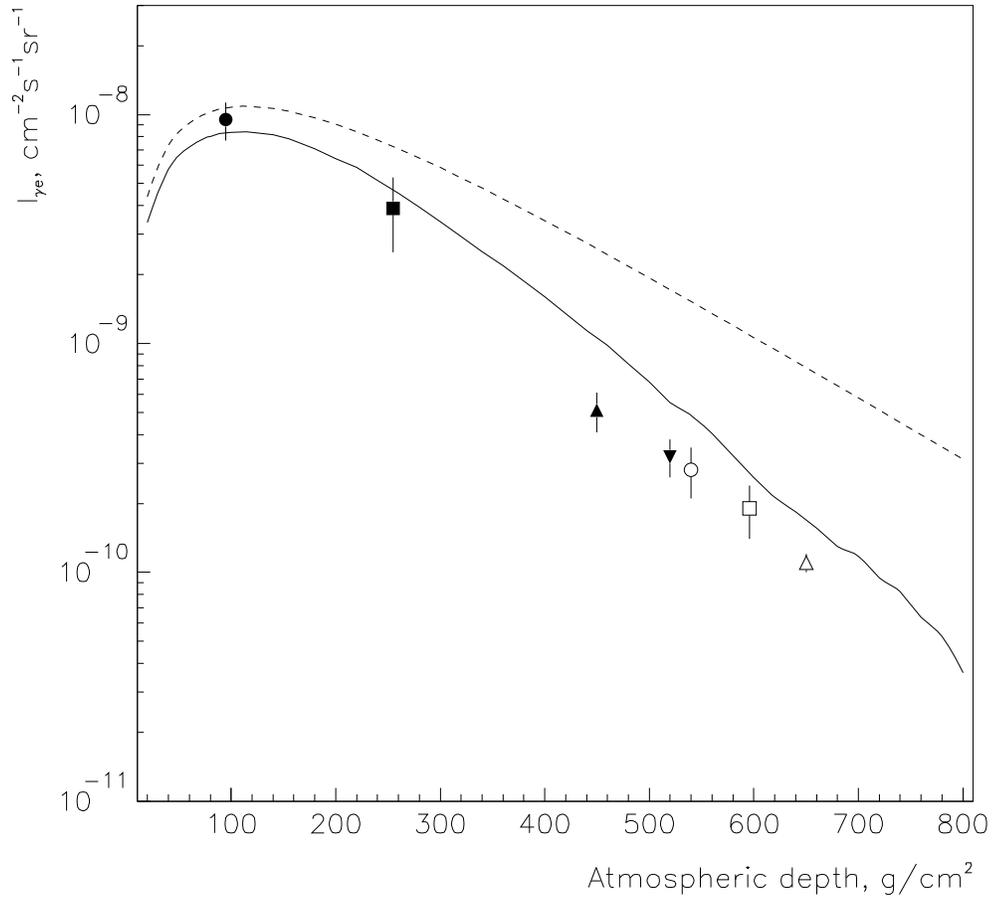,height=15cm}
\caption {Integral flux of electromagnetic component (photons,
electrons, positrons) with energy more than 5 TeV as a function
of depth in atmosphere. Solid curve -- present simulations with
CORSIKA, dashed curve -- parameterisation from \cite{gaisser},
experimental points are the data at different altitudes
(see \cite{gaisser,navia} for references).} \label{fig1}
\end{center}
\end{figure}

We checked our simulations of photons in EAS with CORSIKA against
experimental data at high altitudes.
Simulated flux of electromagnetic component (photons, electrons
and positrons) with energy more than 5 TeV as a function of atmospheric
depth is plotted in Figure \ref{fig1} (solid curve) together
with our calculation using parameterisation given by Gaisser
\cite{gaisser} (dashed line). Figure \ref{fig1} shows that measured
atmospheric depth profile of photon flux agrees better
with our simulations
using CORSIKA. There is still a discrepancy at depths more than
400 g/cm$^2$. More than 90\% of photons
in the showers, however, are above this depth where the agreement
between simulations and measurements is pretty good.

\begin{figure}[htb] 
\begin{center}
\epsfig{figure=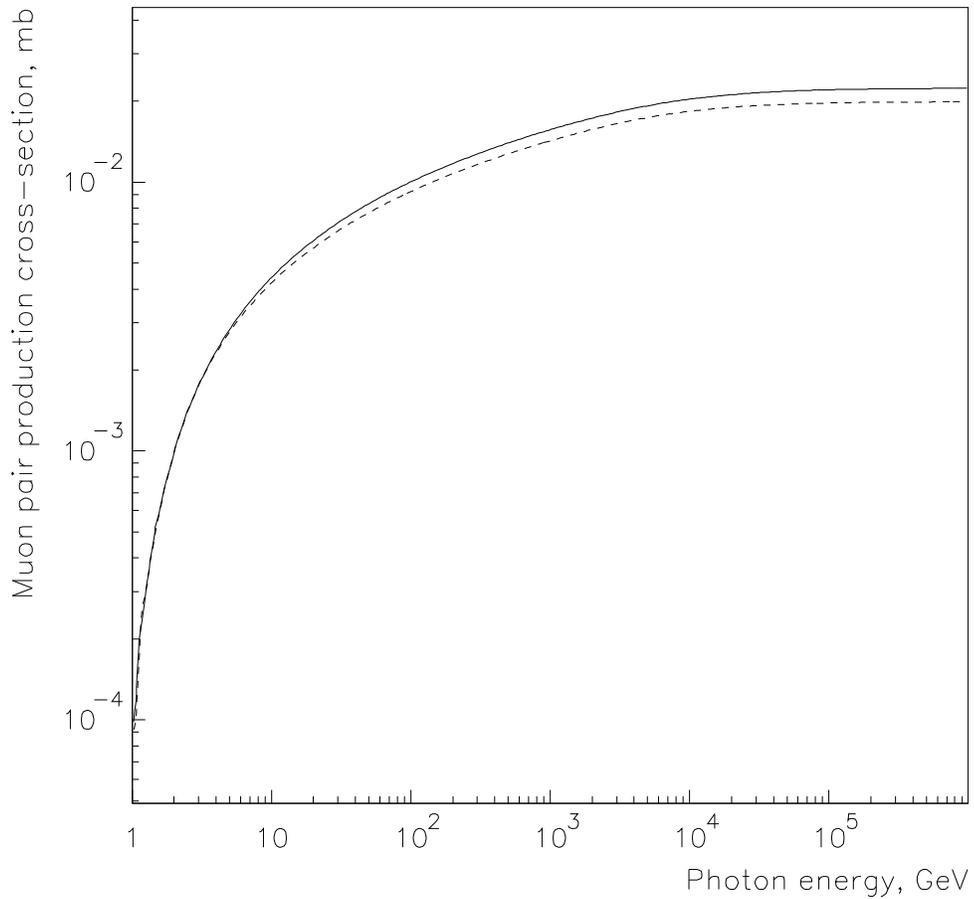,height=15cm}
\caption {Muon pair production cross-section as a function
of photon energy for two minimal values of fractional energy transfer:
$x_{min}=E_{\mu min}/E_{\gamma}=0.01$ (solid curve) and 
$x_{min}=0.1$ (dashed curve).} \label{fig2}
\end{center}
\end{figure}

Muon pair production by photons was simulated using 
differential cross-section for muon bremsstrahlung
\cite{bb}. This cross-section can be easily converted
to muon pair production cross-section by reversing
particles in initial and final states. The muon pair production
cross-section as a function of photon energy is shown in Figure 
\ref{fig2} for two minimal values of fractional energy transfer:
$x_{min}=E_{\mu min}/E_{\gamma}=0.01$ (solid curve) and 
$x_{min}=0.1$ (dashed curve). The cross-section does not change much
with the decrease of the lower limit of integration ($x_{min}$) below
0.01.

Angles of muons
in the final state were sampled according to the method used in GEANT
\cite{geant}. Further transport of muons in atmosphere 
was done using a specially developed version of three-dimensional
muon propagation code MUSIC (standard version of the code is described
in Ref. \cite{music}). We found that muon
deflection at production and multiple scattering in the
atmosphere contribute significantly to the lateral
separation of muons in pairs underground. Muon deflection
at production is important for muon separation at shallow depths 
(low muon energies) since mean deflection angle is
proportional to $m_{\mu}/E_{\mu}$. Note, that since the air 
density varies strongly with atmospheric depth, 
the step along muon path to calculate
deflection due to multiple scattering should be small.
For comparison, we calculated also muon separation underground
without accounting for muon deflection at production and/or
multiple scattering in atmosphere.

Muon transport through rock or water down to the observation level was
done using standard version of muon propagation code MUSIC
\cite{music}. 

\vspace{0.7cm}
\noindent {\large \bf 3. Muon pair production by photons in hadronic 
showers}
\vspace{0.5cm}

Muon pair production is negligible compared to the conventional muon
production at TeV energies. Due to the competition between
interaction and decay for pions and kaons, however, power
index of conventional muon spectrum is about 3.7 while
direct muon energy spectrum is harder with power index 
$\gamma=2.7$. This
implies that at some energy direct muon pair production
may dominate over conventional muons (see \cite{vak2} for
detailed discussion). Our three-dimensional calculations
of vertical EAS initiated by primary protons
show that the ratio of direct muons to pions decreases from
$2.4 \times 10^{-5}$ at 10 TeV down to $1.6 \times 10^{-5}$
at 1000 TeV assuming constant slope of pion spectrum. 
The decrease of the ratio is due to the steepening
of the photon spectrum with energy.
In the simulations we use constant slope of primary
spectrum at all energies. The change of the slope at and above
the ``knee'' will affect absolute values of pion, photon and
muon fluxes but is unlikely to change significantly the
ratio of direct muons to pions and conventional muons.
The calculated ratio of direct muons to pions 
is smaller than that
obtained in Ref. \cite{vak2}, where parameterisation
from Ref. \cite{gaisser} for photon flux as a 
function of energy and depth in atmosphere was used.
The formula in Ref. \cite{gaisser} was obtained assuming 
scaling parameterisation
of the inclusive spectra and constant cross-sections.
It does not fit measured atmospheric profile of
photon flux (see Figure \ref{fig1} for comparison).

We simulated a sample of helium-  and iron-induced showers and 
found that for the same energy per nucleon the numbers of muons
and photons in EAS are proportional to the atomic weight of
primary nucleus. We concluded that
our approach to use primary spectrum of all nucleons is justified
regardless of the precise composition of primary cosmic rays.
If heavy nuclei dominate at energies above $10^{15} - 10^{16}$ eV,
however, mean muon multiplicity per shower will be higher than
for proton primaries.
Muons from direct photoproduction
dominate over conventional muons at vertical at energies
higher than $1.5 \times 10^{16}$ eV (the intersection point at 
$3 \times 10^{15}$ eV was obtained in \cite{vak2}). It is hard
to predict, however, the spectra of conventional muons
and direct muons near and beyond
the region of the ``knee'' in the primary spectrum.
Note, that muons from charm particle decay may dominate in the
total muon flux at these energies since most
models predict the ratio of charm-produced muons
to pions larger than or about $10^{-4}$.

The number of directly photoproduced muons with energy higher 
than $10^3$ TeV is about 5 per year per steradian for 1 km$^2$ 
detector, while corresponding number of conventional muons is 
about 30 at vertical. Both numbers suffers from large uncertainties 
in the primary spectrum, its composition and model of nucleus-nucleus 
interaction at high energies. 

It is unlikely that even future 1 km$^2$ neutrino telescopes,
capable of measuring muon energy, can detect and discriminate directly 
photoproduced muons 
from conventional muons and/or prompt muons (from charmed particle
decay). Although direct muons have a distinctive
feature, such as flat zenith angular distribution (similar to that
of prompt muons), the number of events is too small for this feature
to be seen. At energies about 100 TeV those events are hidden by 
much larger number of conventional muons. At higher energies (more
than 1000 TeV) the number of events is too small and many years
of exposure are needed to collect reasonable statistics. Moreover,
it is likely that prompt muons dominate in the total muon flux at
these energies. It is not clear also how sensitive neutrino telescopes
will be to the down-going muons, how accurately they can measure muon
energy etc.
 
\vspace{0.7cm}
\noindent {\large \bf 4. Muon pair production by photons in 
electromagnetic showers}
\vspace{0.5cm}

To quantify muon pair production by photons in electromagnetic
showers we used the same approach as in Ref. \cite{vs1}. We
characterised muon pair production by its yield, i.e. the
ratio of muon flux above given energy to primary photon flux
above the same energy, $r_{\mu}(>E)=F_{\mu}(>E)/F_{\gamma}(>E)$.

At first we checked the yields of directly photoproduced muons
and conventional muons from the CORSIKA code itself. We found
the yield for directly photoproduced muons to be about
$(3-6) \times 10^{-5}$ for energy $E=1-10$ TeV and primary index 
$\gamma=2.1$, which is a few times
less than obtained in Ref. \cite{vs1}. Much higher statistics
(and more CPU time) is required for better estimate of
the direct muon yield in CORSIKA. Observed
discrepancy supports our decision to use home-made
programs with cross-section valid at high energies
\cite{bb} to simulate muon pair production in atmosphere.

Results of the simulations are shown in Table \ref{table1}
together with earlier calculations from Ref. \cite{vs1}.
$10^6$ showers were simulated for each case giving the
statistics for conventional muon yield 
$r_{\mu}(>E) \times 10^{6}$ (for example, 1802 muons were
obtained for $E=0.1$ TeV and $\gamma=2.1$, which gives the
value of yield $r_{\mu}(>E)=1.8 \times 10^{-3}$ in Table 
\ref{table1}). 
This makes the results for conventional muons quite uncertain
at threshold energies $E>>1$ TeV. However, at such energies
the contribution of this process to the total muon flux is 
quite small. Much higher statistics was achieved for direct muon
production at these energies since the average atmospheric depth
profile of photons was used to simulate this process independently
of the development of a particular shower.
Our simulations are
in reasonable agreement with previous results \cite{vs1}.
Note, that the calculations
in Ref. \cite{vs1} have been performed semi-analytically within
Approximation A of the cascade theory without full Monte Carlo.
The observed difference (less than a factor of 2 for all energies
and spectral indices) can be explained by more 
adequate model of photopion production in CORSIKA and/or
by precise treatment of the development of electromagnetic
shower in atmosphere.

\begin{figure}[htb] 
\begin{center}
\epsfig{figure=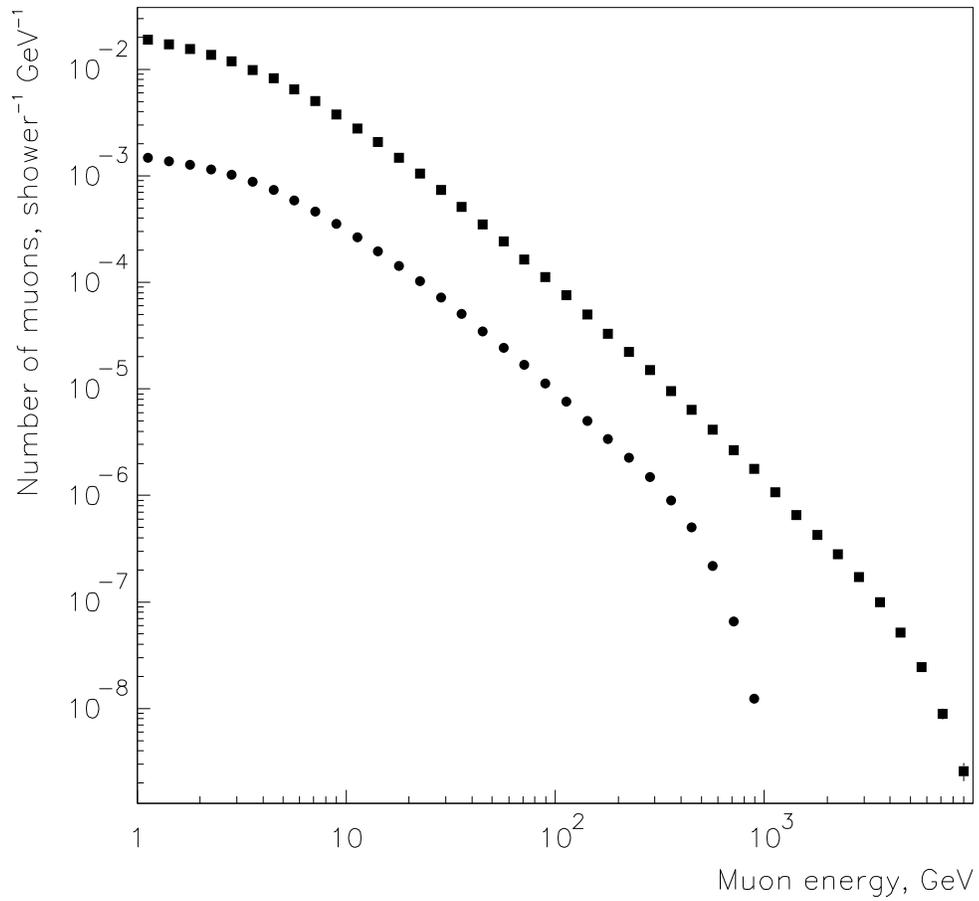,height=15cm}
\caption {Differential muon spectra in electromagnetic showers at sea level
at vertical for 1 TeV (filled circles) and 10 TeV (filled squares)
primary photons.} \label{fig3}
\end{center}
\end{figure}

Differential muon spectra in electromagnetic showers at sea level
at vertical for 1 TeV and 10 TeV primary photon energies are shown 
in Figure \ref{fig3}. The suppression of the spectra at low energies is
due to the muon decay in the atmosphere.

Good agreement between present Monte Carlo simulations and
previous se\-mi-ana\-ly\-tical computations \cite{vs1} in terms of muon 
yields from various processes proves that main conclusions 
of Ref. \cite{vs1} are still valid. Since the end of 1980s, however,
very high energy gamma-ray astronomy made significant progress
(see \cite{ong,catanese,kr} for recent reviews). A number of
galactic and extragalactic sources has been detected at TeV 
energies, the Crab Nebula being the brightest of them.
Typical fluxes from the sources or upper limits from
ground-based observations by existing atmospheric Cherenkov telescopes 
are of the order of $10^{-11}$ photons/(cm$^2$ s) at 1 TeV.
Recent results from ground-based telescopes can be used now
to estimate the sensitivity of underground/underwater/under-ice 
detectors to muons produced in electromagnetic cascades.

We started with the brightest source -- the Crab Nebula. The 
differential energy
spectrum from this source can be approximated as
$3 \times 10^{-11} \times (E/$TeV$)^{-2.5}$ 
photons/(cm$^2$ s TeV) 
(see \cite{catanese} and references therein). 
Assuming no high-energy cut-off of the spectrum,
the flux of directly photoproduced muons from the
Crab Nebula is $9.5 \times 10^{-16}$ cm$^{-2}$ s$^{-1}$
above 1 TeV at surface.
Muon flux from the Crab Nebula was also calculated in Ref.
\cite{bhatt}. Although the author of Ref. \cite{bhatt}
assumed smaller photon flux from the source above 1 TeV
(integral spectrum 
$1.07 \times 10^{-11} \times (E/$TeV$)^{-1.4}$ 
photons/(cm$^2$ s) against
$2.0 \times 10^{-11} \times (E/$TeV$)^{-1.5}$ 
photons/(cm$^2$ s) assumed
in present simulations) the muon flux from direct
pair production quoted in Ref. \cite{bhatt} is several times
higher than our result. To calculate the flux from muon pair 
production Bhattacharyya \cite{bhatt} followed analytical
procedure developed by Berezinsky et al. \cite{vs1}. However,
he used photopion production cross-section 
(0.332 mb, see eq.(8) in Ref. \cite{bhatt}) 
instead of muon pair production cross-section
(asymptotic value $\approx$0.022 mb as in Figure \ref{fig2} and
Ref. \cite{vs1}).

The flux of conventional muons from the Crab Nebula above 1 TeV
at surface is 1.5 times larger 
($1.5 \times 10^{-15}$ cm$^{-2}$ s$^{-1}$) 
than that of direct muons. Again this value is
several times smaller than the flux calculated
in Ref. \cite{bhatt}.

Similar considerations applied to the ``standard'' source
with integral spectrum $10^{-11} \times (E/$TeV$)^{-1.1}$ 
photons/(cm$^2$ s) reveal the following fluxes above 1 TeV:
$1.26 \times 10^{-15}$ cm$^{-2}$ s$^{-1}$ from
direct muon pair production and 
$3.05 \times 10^{-15}$ cm$^{-2}$ s$^{-1}$ from
pion and kaon decay. This corresponds to 1360 muons with
energy at surface above 1 TeV which
can be observed by a detector with effective area of 1 km$^2$
per year. The number of background atmospheric 
muons in a cone with half-angle 
1$^{\circ}$ is $1.6 \times 10^{7}$. This gives a signal-to-noise
ratio $S/\sqrt{N}$=0.34. The ratio decreases with decreasing
energy threshold and detector area.
The high-energy cut at 100 TeV
reduces the number of muons from the source by a factor of 3.
Our results are more pessimistic than calculations by Halzen et al.
\cite{halzen} due to the difference in the photon fluxes
used in the calculations. We used measured photon 
fluxes at about 1 TeV (see, for example, Ref. \cite{catanese} 
and references therein) while Halzen et al. \cite{halzen}
extrapolated photon spectra measured at about 100 MeV to TeV
energies.

\vspace{0.7cm}
\noindent {\large \bf 5. Sensitivity of underground/underwater detectors
to muons from point sources}
\vspace{0.5cm}

Better way to estimate the sensitivity of underground/underwater
detectors to muons from gamma-ray sources is to calculate the
characteristics of muons fluxes at the detector site, namely,
energy spectra, lateral and angular distribution of photoproduced
muons together with muon background. We consider here three types
of underground/underwater detectors: i) existing underground
detectors, like MACRO and LVD (see, \cite{macro,lvd} and references
therein for detector description), 
with good spatial and angular resolution but incapable
of measuring energy of through-going muons, their surface
area is of the order of 1000 m$^2$, ii) shallow depth
underground detectors, like LEP experiments (for Cosmo-LEP
project see \cite{l3,aleph}), with relatively small area (about 
100-200 m$^2$) but
good spatial, angular and energy resolution, and iii) existing
and future underwater/under-ice neutrino telescopes, like
AMANDA \cite{amanda} and ANTARES \cite{antares} with effective
area of 1 km$^2$.

We assumed that detector of type i) is at the depth of 3 km w.e.
of standard rock, detector of type ii) is at the depth of 
0.3 km w.e. of standard rock and detector of type iii) is at
the depth of 2 km of water/ice. We assumed also that integral 
gamma-ray spectrum from ``standard source'' can be approximated as
$10^{-11} \times (E/$TeV$)^{-1.1}$ photons/(cm$^2$ s)
above 100 GeV without high-energy cut-off. The cut-off
at 100 TeV reduces the muon flux above 1 TeV by a factor of 3 if
power index of differential spectrum is $\gamma=2.1$.
If $\gamma=2.5$, the reduction does not exceed 30\%.
We considered only vertical muons and only one depth for each
detector specified
above. Realistic slant depth distribution is important
for any particular experiment but cannot change qualitatively
our results. 

Detector of type i) at 3 km w.e. will detect 0.85 muons
per year per 1000 m$^2$ of effective area. It is obvious
that none of existing or planned underground detectors
is able to discriminate such a tiny flux from the 
background which is of the order
of 6000 muons in a cone with half-angle 1$^{\circ}$
(solid angle of about $10^{-3}$ sr). Number of
double muon events from direct muon pair production 
is 0.054 with additional (50-70)\% contribution from
pion and kaon decays. Average lateral separation of direct
muons in pairs is of the order of 3 meters, which
is less than typical separation of muons in muon
bundles originated in hadronic EAS. There is no way, however,
to use this feature for definite detection of muons from
the source.

At the depth of 0.3 km w.e. muons from pion and kaon decays
dominate and the contribution of direct muons is negligible. 
Number of muons from the ``standard source'' which 
can be detected by 100 m$^2$ detector is about 17 per year.
The muon background from hadronic showers is about $6.8 \times 10^5$
in a cone with half-angle 1$^{\circ}$.
Again, the statistics for the signal is too small for
positive detection of the source. The number of double muon
events from both direct muon production and conventional
muon source is about 1 per year per 100 m$^2$ which is
much smaller than the number of background double muon events
initiated in hadronic EAS. The spread of muons in double muon
events is large (several ten meters for both signal and background).
Detectors of type ii) are able to measure muon energy using
magnets. With increasing energy threshold the signal-to-noise
ratio increases, in particular, for double muon events, but
this does not help to detect the signal because of significant
reduction in the number of expected events.
In the simulations we neglected the decay of directly produced muons 
which will have only minor effect on muon intensities above 50 GeV at 
vertical.

Finally, at the depth of 2 km in water or ice 1 km$^2$ detector
will detect about 2500 muons with energy higher than 100 GeV
(typical energy threshold for through-going muons for 
such a detector) per year. The number
of background muons is about $1.4 \times 10^7$ muons per
year in an 1$^{\circ}$ half-angle cone. The signal-to-noise ratio is
about 0.67 which is not enough for positive detection of the
signal from the source. Moreover, part of muons from the source
are coming in groups which can be seen as single muons, thus
reducing the total number of detected events. Assuming that such a
detector can estimate muon energy, the software energy
threshold of 1 TeV for through-going muons can be applied 
but is unlikely to change the
situation in favour of the signal. The number of background muons
in this case is $1.1 \times 10^6$, while about 540 muons from the
source can be detected with signal-to-noise ratio of about 0.5.
The use of double muon events with small separation between muons
as a signature for muon pair production is doubtful because
it is not known at present how such a detector can discriminate
multiple muon events from single muons.

Muon production in electromagnetic showers has been recently considered
by Fass\`o and Poirier \cite{poirier1} and Poirier et al. 
\cite{poirier2}. These authors presented detailed simulations of muons
in electromagnetic showers using FLUKA Monte Carlo code \cite{fluka}
(without muon pair production) in application to the GRAND \cite{grand}
and Milagro \cite{milagro} experiments. They concentrated on the
characteristics of muon flux but did not discuss the sensitivity
of detectors to muons from astrophysical gamma-ray sources.
Conventional muon yields obtained in present work with CORSIKA 
(QGS model with jets) are in
reasonable agreement with the results from FLUKA \cite{poirier1}.
For 1 TeV and 10 TeV primary photons muon spectra from CORSIKA are about
20\% lower than corresponding results from FLUKA \cite{poirier1}.
Extrapolation of our results from $\approx$100 GeV down to $\approx$1
GeV (for signal and background muon spectra) shows that the signal-to-noise
ratio remains much less than 1. It is not obvious, however, that such
an extrapolation is a correct procedure for photon spectra from
point sources since they have not been measured directly at 1-500 GeV.
There is still a possibility that at GeV energies photon fluxes are
higher than expected from such a simple extrapolation. Only future
measurements of (or upper limits on) the fluxes of photons or muons
at these energies will provide necessary information.

The possibility to use muons as detectors of gamma-ray bursts was discussed
by Halzen et al. \cite{halzen} and Alvarez-Mu\~niz and Halzen \cite{halzen1},
but is beyond the scope of present study.

\vspace{0.7cm}
\noindent {\large \bf 6. Conclusions}
\vspace{0.5cm}

The calculations presented here were based on the measured very high energy
gamma-ray fluxes from a number of sources and the limits on
the fluxes from some other point sources. Our simulations show
that the expected high-energy muon fluxes under\-gro\-und/\-un\-der\-water/un\-der-ice
are too small (also in comparison with muon background)
to be detected by existing and planned
detectors including large-scale neutrino telescopes.
Specific signatures of high-energy muon events initiated
by photons (muon pairs
with small lateral separation of muons) does not help
to detect signal.
High-energy muon astronomy can survive if
i) extremely powerful steady, pulsed or (most likely) 
burst sources of very high energy photons exists in the Universe;
ii) a new channel of muon production is found or the cross-section(s)
of the process(es) which lead(s) to muon production at high
energies is (are) much higher than expected;
iii) a new neutral, stable, strongly interacting
particle substitutes photon in muon production.
However, only continuous monitoring of the sky can help to answer
these questions.

\vspace{0.7cm}
\noindent {\large \bf Acknowledgments}
\vspace{0.5cm}

\indent The author is grateful to Prof. V. S. Berezinsky for stimulation
of the work and very fruitful discussions. I wish to thank Dr.
E. V. Korolkova for useful comments. I appreciate also the remarks from
anonymous referee.

\pagebreak

\begin{table}[htb]
\caption{ Muon yields (ratio of muon flux to primary
photon flux above the same energy $r(E)=F_{\mu}(>E)/F_{\gamma}(>E)$) 
multiplied by $10^6$ for various energies $E$ and indices of primary
spectrum $\gamma$. The row $\gamma \rightarrow 2 \mu$ shows the
yield of direct muon pair production, $\pi, K \rightarrow \mu$ shows
contribution from conventional muons.} \label{table1}
\vspace{1cm}
\begin{center}
\begin{tabular}{|c|c|cccccc|}\hline
          & $\gamma$              & 2.1 & 2.1 & 2.1 & 2.5 & 2.5 & 2.5 \\ \hline
          & $E$, TeV                & 0.1 & 1   & 10  & 0.1 & 1  & 10 \\ \hline
          & $\gamma \rightarrow 2\mu$ & 95 & 127 & 126 & 32 & 47 & 56 \\  
This work & $\pi, K \rightarrow \mu$ & 1802 & 305 & 35 & 386 & 76 & 9 \\  
          & sum                   & 1897 & 432 & 161 & 418 & 123 & 65 \\ \hline 
          & $\gamma \rightarrow 2\mu$ & & 156 & 181 & & 25.5 & 30.4  \\  
\cite{vs1} & $\pi, K \rightarrow \mu$ & & 358 & 40  & & 42.0 & 4.7   \\  
           & sum                       & & 514 & 221 & & 67.5 & 35.1  \\ \hline 
\end{tabular}
\end{center}
\end{table}

\pagebreak

\pagebreak

\pagebreak

\end{document}